\documentclass[aps,twocolumn,pra,showpacs,floatfix]{revtex4}
\usepackage{epsfig}
\usepackage{graphicx}
\usepackage{dcolumn}
\usepackage{amssymb,amsmath}
\usepackage{mathrsfs}
\begin{document}

\title{Calculation of hyperfine structure of erbium and fermium.}

\author{V. A. Dzuba, V. V. Flambaum}

\affiliation{School of Physics, University of New South Wales, Sydney 2052, Australia}

\begin{abstract}
A version of the configuration interaction method, which has been recently developed to deal with large number of valence electrons, has been used to calculate magnetic dipole and electric quadrupole hyperfine structure constants for a number of states of erbium and fermium.
Calculations for fermium are done for extracting nuclear moments of Fm isotopes from recent and future measurements.
Calculations for erbium, which has electronic structure similar to those of fermium, are done to study the accuracy of the method.

\end{abstract}


\maketitle

\section{Introduction}

Spectroscopic study of heavy actinides have shown good progress in recent years~\cite{Fm1,Fm2,ErFm,Es2009,Raeder2022,Mustapha2022,Raeder,CfEs}.
A particular focus of this study was on the hyperfine structure (hfs). 
Comparing measured and calculated hfs leads to extraction of nuclear moments advancing our knowledge on nuclear structure of heavy elements.
This in turn may benefit the search for the hypothetical stability island, i.e.  superheavy nuclei with a long lifetime. There is strong correlation between the value of the  electric quadrupole moment $Q$ and nuclear deformation. Larger deformation usually means larger value of $Q$. 
On the other hand, the nuclei in the vicinity of the stability island are expected to be spherical.
 Therefore, observing elements with small $Q$ may indicate approaching the stability island. 

Hyperfine structure of $^{255}$Fm~\cite{Fm2},  $^{254}$Es~\cite{Es2009}, $^{253-255}$Es~\cite{Raeder2022}, $^{249-253}$Cf~\cite{Raeder} have been  measured, and comparison with calculations~\cite{ErFm,Raeder,CfEs} leads to determination of the magnetic dipole ($\mu$) and electric quadrupole ($Q$) nuclear moments of corresponding isotopes of Es and Cf. For these atoms we calculated hfs in the ground state only~\cite{CfEs}. In principle, this is sufficient to determine nuclear moments.
However, the situation is more complicated for $^{255}$Fm isotope. 
Experimental paper~\cite{Fm2} gives two conflicting interpretations of the hfs splitting in the ground and two excited states. 
Calculations of hfs for all these three states~\cite{ErFm} did not resolve the problem. New measurements are currently in progress~\cite{SRaeder}.
In this paper we present more detailed and accurate calculations for the hfs of Fm in hope to assist in the interpretation of the experimental data.
The calculations include a number of excited states which are connected to the ground state via electric dipole transitions. The energies of these states were calculated in our previous paper~\cite{ErFm}. Seven of the energy levels were measured experimentally~\cite{Fm1,Fm2}.
In the present paper we calculate hfs for most of these states.
We also calculate hfs of Er, which is lighter analog of Fm, to assess the accuracy of the calculations. 

\section{Method of calculations}

In this paper we mostly follow our previous work on Dy, Ho, Cf and Es~\cite{CfEs}. Calculations of the energies and wave functions are performed with the use of the CIPT (configuration interaction with perturbation theory) method~\cite{cipt}. This method was specially developed for open-shell atoms with a large number of valence electrons. 
Er and Fm have fourteen valence electrons each (the $4f^{12}6s^2$ ground state configuration of external electrons in Er and the $5f^{12}7s^2$ ground state configuration in Fm). 
The basis of many-electron single-determinant wave functions for fourteen electrons is divided into two parts: low energy states and high energy states.
External electron wave functions are expressed in terms of coefficients of expansion over single-determinant basis state functions 
\begin{eqnarray}
&&\Psi (r_1, \dots ,r_M)=   \label{e:Psi}  \\
&&\sum_{i=1}^{N_1} x_i \Phi_i (r_1, \dots ,r_M)+ \sum_{j=1}^{N_2} y_j \Phi_j(r_1, \dots ,r_M). \nonumber
\end{eqnarray}
Here $M$ is the number of valence electrons.
{ The terms in (\ref{e:Psi}) are ordered according to the energies of the  single-determinant functions, from low to high energies, 
$\langle \Phi_{I-1}|\hat H^{\rm CI}|\Phi_{i-1}\rangle < \langle \Phi_i|\hat H^{\rm CI}|\Phi_i\rangle < \langle \Phi_{i+1}|\hat H^{\rm CI}|\Phi_{i+1}\rangle$. }
 $N_1$ is the number of low-energy basis states, $N_2$ is the number of high-energy basis states.
It is assumed that $N_1 \ll N_2$ and that first $N_1$ terms in (\ref{e:Psi}) represent good approximation for the wave function while the rest of the sum is just small correction.
Then the CI matrix equation can be written in a block form
\begin{equation} \label{e:blocks}
\left( \begin{array}{cc} \mathcal{A} & \mathcal{B} \\ \mathcal{C} & \mathcal{D} \end{array} \right) \left(\begin{array}{c} \mathcal{X} \\ \mathcal{Y} \end{array} \right) = E_a \left(\begin{array}{c} \mathcal{X} \\ \mathcal{Y} \end{array} \right).
\end{equation}
Here block $\mathcal{A}$ corresponds to low-energy states, block $\mathcal{D}$ corresponds to high-energy states, and~blocks $\mathcal{B}$ and $\mathcal{C}$ correspond to cross terms. 
Note that since the total CI matrix is symmetric, we have $\mathcal{C} = \mathcal{B}^t$, i.e.,~$c_{ij} = b_{ji}$. 
Vectors $\mathcal{X}$ and $\mathcal{Y}$ contain the coefficients of expansion of the valence wave function over the single-determinant many-electron basis functions (see Eq.~\ref{e:Psi}).

The main feature of the CIPT method~\cite{cipt} is neglecting the off-diagonal matrix elements in block $\mathcal{D}$. This allows one to greatly simplify the CI equations (\ref{e:blocks}) reducing the size of the CI matrix to the size of block $\mathcal{A}$ (see Ref.~\cite{cipt} for details).

Finding $\mathcal{Y}$ from the second equation of (\ref{e:blocks}) leads to
\begin{equation}\label{e:Y}
\mathcal{Y}=(E_aI-\mathcal{D})^{-1}\mathcal{CX}.
\end{equation}
Substituting $\mathcal{Y}$ to the first equation of (\ref{e:blocks}) leads to
\begin{equation}\label{e:CIPT}
\left[\mathcal{A + B}(E_aI-\mathcal{D})^{-1}\mathcal{C}\right] \mathcal{X} = E_a \mathcal{X},
\end{equation}
where $I$ is  the unit matrix. 
Then, following Ref.~\cite{cipt} we neglect off-diagonal matrix elements in block $\mathcal{D}$. This leads to a very simple structure of the $(E_aI-\mathcal{D})^{-1}$ matrix, $(E_aI-\mathcal{D})^{-1}_{ik} = \delta_{ik}/(E_a - E_k)$, where $E_k = \langle k|H^{\rm CI} |k \rangle$.
Note that unknown energy of the state of interest $E_a$ can be found in both right and left-hand side of (\ref{e:CIPT}). 
This means that iterations over $E_a$ are needed to solve (\ref{e:CIPT}). Initial approximation for $E_a$ can be found from solving $\mathcal{AX} = E_a\mathcal{X}$.

\begin{table}
\caption{\label{t:N1N2}
Typical number of the dominating terms in the wave function expansion (\ref{e:Psi}) ($N_1$, which is equal to the size of the effective CI matrix) and the number of terms in the correction ($N_2$), for the ground and exited odd states of Er and Fm.}
\begin{ruledtabular}
\begin{tabular}   {l c rr}
\multicolumn{1}{c}{States}&
\multicolumn{1}{c}{$J$}&
\multicolumn{1}{c}{$N_1$}&
\multicolumn{1}{c}{$N_2$}\\
\hline
Ground state & 6 & 2 & $\sim 8.1 \times 10^6$ \\
Odd states    & 5 & 74 & $\sim 3.4 \times 10^8$ \\
                     & 6 & 58 & $\sim 1.0 \times 10^8$ \\
                     & 7 & 38 & $\sim 2.7 \times 10^8$ \\
\end{tabular}			
\end{ruledtabular}
\end{table}

{ Typical values of $N_1$ and $N_2$ for Er and Fm are presented in Table~\ref{t:N1N2}. The values of $N_1$ correspond to the minimal option, in which dominating terms for the ground state are represented only by the states of the $4f^{12}6s^2$ configuration for Er and $5f^{12}7s^2$ configuration for Fm, while for excited odd states dominating terms include states of two odd configurations, the $4f^{12}6s6p$ and $4f^{11}6s^25d$ configurations for Er and the $5f^{12}7s7p$ and $5f^{11}7s^26d$ configurations for Fm. 
For the minimal option the calculations can be done on a laptop or similar computer. In principle, one can try to improve the accuracy of the calculations by including more terms into the low-energy part of the expansion (\ref{e:Psi}). However, this is a computationally expensive path.
The computational time is roughly proportional to  $N_1 \times N_2$ since most of time goes to calculation of the second-order correction to the effective CI matrix (second term in the left-hand side of Eq.~(\ref{e:CIPT}), which is a rectangular matrix of the $N_1 \times N_2$ size). 
On the other hand, there is usually significant energy gap between the states of the lowest and excited configurations of an atom. This means that moving just few terms from the second to the first part of expansion (\ref{e:Psi}) would not change the result much. One has to increase the value of $N_1$ significantly to see any real change. This may lead to significant increase of the computational time.}

To calculate hfs, we use the time-dependent Hartree-Fock (TDHF) method~\cite{CPM,DFS84}, which is equivalent to the well-known random-phase approximation (RPA). { The TDHF method deals with oscillating external fields. The case of hfs corresponds to zero frequency of oscillations, so no real time dependence is introduced. }
The RPA equations can be written as:	
\begin{equation}\label{e:RPA}
	\left(\hat H^{\rm RHF}-\epsilon_c\right)\delta\psi_c=-\left(\hat f+\delta V^{f}_{\rm core}\right)\psi_c
\end{equation}
where $\hat H^{\rm RHF}$ is the relativistic Hartree-Fock Hamiltonian, $\hat f$ is an operator of  an external field (nuclear magnetic dipole or electric quadrupole fields).  This operator takes into account finite nuclear size for both, magnetic dipole~\cite{DFS84} and electric quadrupole~\cite{FNS} operators.

Index $c$ in (\ref{e:RPA}) numerates states in the core, $\psi_c$ is a single-electron wave function of the state $c$ in the core, 
{ $\epsilon_c$ is its Hartree-Fock energy,}
$\delta\psi_c$ is the correction to this wave function caused by an external field, and $\delta V^{f}_{\rm core}$ is the correction to the self-consistent RHF potential caused by changing of all core states. 
Eq. (\ref{e:RPA}) are solved self-consistently for all states in the core. As a result, the effective operator of the interaction of valence electrons with an external field is constructed as $\hat f + \delta V^{f}_{\rm core}$. The energy shift of a many-electron state $a$ is given by
\begin{equation} \label{e:de}
\delta \epsilon_a = \langle a | \sum_{i=1}^M \left(\hat f+\delta V^f_{\rm core} \right)_i | a\rangle.
\end{equation}
Here $M$ is the number of valence electrons.

When the wave function for the valence electrons comes as a solution of Eq.~(\ref{e:CIPT}), Eq.~(\ref{e:de}) is reduced to
\begin{equation}\label{e:mex}
\delta \epsilon_a = \sum_{ij} x_i x_j \langle \Phi_i|\hat H^{\rm hfs}|\Phi_j \rangle,
\end{equation}
where $\hat H^{\rm hfs} =  \sum_{i=1}^M (\hat f+\delta V^f_{\rm core})_i$.
For better accuracy of the results, the full expansion (\ref{e:Psi}) might be used. Then it is convenient to introduce  a new vector $\mathcal{Z}$, which contains both $\mathcal{X}$ and $\mathcal{Y}$, $\mathcal{Z} \equiv \{\mathcal{X,Y}\}$. Note that the solution of (\ref{e:CIPT}) is normalized by the condition $\sum_i x_i^2=1$. The normalization condition for the total wave function (\ref{e:Psi}) is different,  $\sum_i x_i^2+\sum_j y_j^2 \equiv \sum_i z_i^2=1$. Therefore, when $\mathcal{X}$ is found from (\ref{e:CIPT}), and $\mathcal{Y}$ is found from (\ref{e:Y}), both vectors should be renormalized. Then the HFS matrix element is given by the expression, which is similar to (\ref{e:mex}) but has much more terms 
\begin{equation}\label{e:mez}
\delta \epsilon_a = \sum_{ij} z_i z_j \langle \Phi_i|\hat H^{\rm hfs}|\Phi_j \rangle.
\end{equation}
Energy shift (\ref{e:de}) is used to calculate hfs constants $A$ and $B$ using textbook formulas
\begin{equation}
A_a = \frac{g_I \delta \epsilon_a^{(A)}}{\sqrt{J_a(J_a+1)(2J_a+1)}},
\label{e:Ahfs}
\end{equation}
and
\begin{equation}
B_a = -2Q \delta \epsilon_a^{(B)}\sqrt{\frac{J_a(2J_a-1)}{(2J_a+3)(2J_a+1)(J_a+1)}}. 
\label{e:Bhfs}
\end{equation}
Here $\delta \epsilon_a^{(A)}$ is the energy shift (\ref{e:de})  caused by the interaction of atomic electrons with the nuclear magnetic moment $\mu$, $g_I=\mu/I$, $I$ is nuclear spin; $\delta \epsilon_a^{(B)}$ is the energy shift (\ref{e:de}) caused by the interaction of atomic electrons with the nuclear electric quadrupole moment $Q$ ($Q$ in (\ref{e:Bhfs}) is measured in barns). 

\vspace{5mm}

The uncertainty of the hfs calculations comes from two sources. One is the uncertainty in the wave function and another one is the contribution of omitted terms in the correlation corrections to the hfs operator. The uncertainty in the wave function is mostly due to limitations of the basis and the fact that most of the mixing states are treated perturbatively. Corresponding effect on the hfs constants ranges from few per cent for large constants to $\sim$ 50\% for small constants. The latter is because small value of the hfs constant comes as a result of strong cancellation between different contributions. Such cancellation leads to the loss in accuracy.
In present calculations we neglect some minor contributions to the correlation corrections to the hfs operator, such as structure radiation, self-energy correction, renormalisation of the wave function and two-particle correction (see, e.g. \cite{hfs-operator}). The combined effect of such corrections does not exceed 10\%~\cite{hfs-operator}. In the end we conclude that the expected accuracy of the hfs calculations is about 10\% for large hfs constants and $\sim$ 50\% for small hfs constants.

\section{Hyperfine structure of erbium}

\begin{table*}
\caption{\label{t:Erhfs}
Energy levels and hyperfine structure constants $A$ and $B$ for low states of  $^{167}$Er. Nuclear spin $I=7/2$, nuclear magnetic moment $\mu(^{167}{\rm Er})=-0.56385(12)\mu_N$~\cite{Stone};
nuclear electric quadrupole moment $Q(^{167}{\rm Er})=3.57(3)~b$~\cite{Stone}; $g_I=\mu/I$.
Last column gives references to experimental data.} 
\begin{ruledtabular}
\begin{tabular}   {ll rr drr drrr}
\multicolumn{1}{c}{Configu-}&
\multicolumn{1}{l}{Term}&
\multicolumn{2}{c}{Energy (cm$^{-1}$)}&
\multicolumn{3}{c}{$A$ [MHz] }&
\multicolumn{3}{c}{$B$ [MHz] }&
\multicolumn{1}{c}{Ref.}\\
\multicolumn{1}{c}{ration}&
&\multicolumn{1}{c}{NIST\cite{NIST}}&
\multicolumn{1}{c}{CIPT}&
\multicolumn{1}{c}{Expt.}&
\multicolumn{1}{c}{present}&
\multicolumn{1}{c}{\cite{ErFm}}&
\multicolumn{1}{c}{Expt.}&
\multicolumn{1}{c}{present}&
\multicolumn{1}{c}{\cite{ErFm}}& \\

\hline

$4f^{12}6s^2$   & $^3$H$_6$        &     0 &     0 & -120.487& -122 & -117 &-4552.984 & -4880 & -5037 & \cite{Childs83} \\
                & $^3$F$_4$        &  5035 &  5370 & -121.9  & -125 & -122 &   516    &   124 &  1050 & \cite{Childs83} \\
                & $^3$H$_5$        &  6958 &  7244 & -159.4  & -159 & -158 & -4120    & -4566 & -4539 & \cite{Childs83} \\
                & $^3$H$_4$        & 10750 & 10838 & -173.4  & -173 & -174 & -2429    & -2470 & -2600 & \cite{Childs83} \\
                & $^3$F$_3$        & 12377 & 13322 & -143.4  & -142 & -139 &  1236    &  1685 &  1767 & \cite{Childs83} \\
                & $^3$F$_2$        & 13097 & 14599 & -167.2  & -166 & -172 &  1688    &  1828 &  1874 & \cite{Childs83} \\
	        &   	           &       &       &         &      &      &          &       &       &               \\
$4f^{11}5d6s^2$ & $(15/2,3/2)^o_6$ &  7176 &  5449 & -139.957& -142 & -135 &  -709.396& -1092 & -1655 & \cite{Childs86} \\
                & $(15/2,3/2)^o_7$ &  7696 &  6024 & -125.851& -123 & -114 & -3046.052& -2285 & -2230 & \cite{Childs86} \\
                & $(15/2,3/2)^o_8$ &  9350 &  6746 & -119.870& -115 & -104 & -3062.704& -2355 & -2372 & \cite{Childs86} \\
                & $(15/2,3/2)^o_9$ &  8620 &  6152 & -113.582& -110 &  -99 &  -782.987& -1121 & -1733 & \cite{Childs86} \\
                & $(15/2,3/2)^o_7$ & 11888 &  8810 & -126.56 & -130 &      &  -2969   & -2121 &       & \cite{Lipert} \\
	        &                  &       &       &         &      &      &          &       &       &                 \\
$4f^{12}6s6p$   & $(6,1)^o_7$      & 17157 & 17399 & -172.5  & -173 & -173 & -4440    & -4377 & -4391 & \cite{Childs83} \\
\end{tabular}			
\end{ruledtabular}
\end{table*}


The results of calculations of energy levels and magnetic dipole hfs constant $A$ and electric quadrupole hfs constant $B$ for $^{167}$Er are presented in Table~\ref{t:Erhfs} and compared with experiment~\cite{NIST,Childs83,Childs86,Lipert} and with our previous calculations~\cite{ErFm}.
Note that the accuracy for the hfs is generally better than for the energies. This is because experimental energies are given as excitation energies form the ground state. In calculations they are given by the difference $\langle \Psi_i|H^{\rm CI}|\Psi_i\rangle - \langle \Psi_0|H^{\rm CI}|\Psi_0\rangle$, where each wave function $\Psi_i$ for excited state and $\Psi_0$ for the ground state has fourteen electrons and the difference is just small fraction of a per cent of each energy. Strong cancellation between two energies leads to some loss of accuracy. On the other hand, the hfs is given just by the expectation value of the hfs operator $\langle \Psi_i|H^{\rm hfs}|\Psi_i\rangle$. 

There are calculations of the hfs of Er using multiconfiguration Dirac-Fock method (MCDF)~\cite{Cheng,Childs86}.
Our results for the magnetic dipole hfs constant $A$ are significantly closer to the experiment in all cases except one. For odd state at $E=7176 \ {\rm cm}^{-1}$ our result is 1.4\% above the experiment while MCDF calculations give a result which is within 1\% of the experiment~\cite{Childs86}. Four electric quadrupole hfs constants $B$ were considered for even states of Er in Ref.~\cite{Cheng}. For two of them the results of the MCDF calculations are closer to experiment, while for other two our results are closer to experiment. Among these four states the most important one in the content of present paper is obviously the ground state. For the ground state our result for $B$ is 7\% larger than experimental value while the MCDF calculations~\cite{Cheng} give the value which is 2.2\% larger than experiment.  In the end we can conclude that in terms of accuracy of the results our method is similar or better than the MCDF method.


Our previous calculations used only dominating terms in the wave function expansion (formula (\ref{e:mex})), while in the present calculations we use complete expansion (formula (\ref{e:mez})). Comparing the results (see Table~\ref{t:Erhfs}) shows systematic but not always significant improvement in accuracy. Accuracy is good for the ground state; it is $\sim $1\% for magnetic dipole constant $A$ and $\sim $7\% for the electric quadrupole constant $Q$. This is mostly due to simple electronic structure of the ground state and significant separation of it from the states of the same parity and $J$. 
This is general trend for many atoms. 
For this reason we have calculated in Ref.~\cite{CfEs} the hfs of Cf and Es in the ground state only.
However, in the present work we calculate  hfs for excited states as well. As one can see from Table~\ref{t:Erhfs}, the accuracy is good for $A$ constant. It is 1-2\% for states of the $4f^{12}6s^2$ and $4f^{12}6s6p$  configurations and 2-4\% for states of the $4f^{11}5d6s^2$ configuration.

The situation is more complicated for the electric quadrupole hfs constant $Q$. For most of the states of the $4f^{12}6s^2$ configuration the relative difference between theory and experiment is $<10$\%. For two states,  $^3$F$_4$ and  $^3$F$_3$, the accuracy is poor.
This can be probably explained by the fact that the values of $B$ for these states are relatively small, which is the results of cancellation between different contribution. Such cancellation usually leads to poor accuracy. Overall, the accuracy for $B$ is lower than for $A$. 
This is partly due to the sensitivity of $B$ constants to the $s-d$ mixing~\cite{FNS}.

The accuracy for $B$ is significantly lower for the states of the $4f^{11}5d6s^2$ configuration (see Table~\ref{t:Erhfs}). It ranges from -25\% to +50\%. This is probably due to the sensitivity of the $B$ constant to the mixing of the states of two different configurations, the $4f^{12}6s6p$ and the $4f^{11}5d6s^2$ configurations. The mixing is roughly proportional to $\langle 4f6p|r_</r^2_>|5d6s\rangle/\Delta E$ ($r_< = \min(r_1,r_2)$, $r_> = \max(r_1,r_2)$).
The dipole Coulomb integral is large and the energy interval is often small, which means large mixing. On the other hand, matrix elements of the $\hat Q$ operator are 2 to 3 times smaller for the states of the  $4f^{11}5d6s^2$ configuration than for the states of the $4f^{12}6s6p$ configuration.
This means that wrong mixing coefficients (e.g., due to inaccurate value of $\Delta E$) leads to the wrong value of $B$. Note that the values of $A$ are much less sensitive to this mixing due to significantly smaller difference in the values of the matrix elements for the states of these two configurations.

In the end we can conclude that the best accuracy is for the ground state (see also Ref.~\cite{CfEs}). It is $\sim 2$\% for $A$ and $\sim 7$\% for $B$. Among excited states the best accuracy should be expected for those states of the $4f^{12}6s6p$ configuration which are well separated on the energy scale from the states of the $4f^{11}5d6s^2$ configuration and give large values of $A$ and $B$.

\section{Hyperfine structure of fermium}

\begin{table}
  \caption{\label{t:Fm}
    Energy levels, magnetic dipole ($A$) and electric quadrupole ($B$) hyperfine structure constants of the ground state of Fm and odd excited states connected to the ground state by electric dipole transitions. Calculated and experimental energies (in cm$^{-1}$), and calculated $g$-factors are included.
    Letters S, P, D in the first column indicate dominating configurations, $5f^{12}7s^2$, $5f^{12}7s7p$ and $5f^{11}7s^26d$ respectively.} 
\begin{ruledtabular}
\begin{tabular}{rrll rr}
&\multicolumn{2}{c}{CIPT}&
\multicolumn{1}{c}{Experimental}&
\multicolumn{1}{c}{$A/g_I$} &
\multicolumn{1}{c}{$B/Q$} \\
&\multicolumn{1}{c}{Energy}&
\multicolumn{1}{c}{$g$-factor}&
\multicolumn{1}{c}{energy~\cite{Fm1,Fm2}}&
\multicolumn{1}{c}{MHz}&
\multicolumn{1}{c}{MHz}\\
\hline
\multicolumn{6}{c}{Ground state, $J=6$} \\
S &      0 & 1.1619 &             &   759  & -1844 \\
\multicolumn{6}{c}{Odd states with $J=5$} \\
P &  20844 & 1.1507 &              & -1317 &  -1438 \\
D &  23663 & 1.1793 &              &  2094 &   -542 \\
P &  24490 & 1.1967 &              &  2285 &   -543 \\
P &  25542 & 1.1211 & 25111.8(0.2)\footnotemark[1] &   399 &  -1850 \\ 
P &  28497 & 1.1778 & 27389(1.5)   &  1446 &  -1311 \\ 
P &  28779 & 1.2342 & 28185(1.5)   &  2792 &    94 \\ 
D &  30356 & 1.1219 &              &   649 &    632 \\


\multicolumn{6}{c}{Odd states with $J=6$} \\

P &  19023 & 1.2565 &                      & 2319 & -1423  \\
D &  19349 & 1.2853 &                      &  908 &  -601  \\
P &  20229 & 1.0876 &                      & -454 & -1103  \\
D &  24408 & 1.1644 &                      &  835 &  -1100  \\ 
P &  25468 & 1.1856 & 25099.8(0.2)\footnotemark[2] & 1788 & -2008  \\ 
P &  28427 & 1.2459 & 27466(1.5)    & 2588 &   342  \\ 
P &  29072 & 1.1761 & 28377(1.5)    &  637 & -1567  \\


\multicolumn{6}{c}{Odd states with $J=7$} \\

D &  19901 & 1.2373 &                    &  786  & -1262  \\
P &  20409 & 1.1922 &                     & 2733  & -1949  \\
D &  24025 & 1.1528 &                     &  764  &  -1144  \\
P &  25220 & 1.2350 &                     & 2231  & -1817  \\
P &  29367 & 1.1456 & 28391(1.5)   & -232  & -1284  \\  
D &  32668 & 1.1244 &                    &  549  &   346 \\
D &  33273 & 1.0677 &                    &  626  &   484  \\


\end{tabular}
\footnotetext[1]{State R2 in Ref.~\cite{Fm2}.}
\footnotetext[2]{State R1 in Ref.~\cite{Fm2}.}
\end{ruledtabular}
\end{table}


The results of calculations for Fm are shown in Table~\ref{t:Fm}. 
The accuracy of the results, { in terms of expected deviation from experiment} is expected to be very similar to those of Er (see previous section for detailed discussion).
The best accuracy is for the ground state. Among excited states, the best accuracy for the hfs constants $A$ and $B$ should be expected for the states of the $5f^{12}7s7p$ configuration, where the value of these constants is relatively large.
In experimental work~\cite{Fm2} the hfs was measured for the ground and two excited states. The first of these two states, called R1, has the energy $E=25099.8(2) \ {\rm cm}^{-1}$, the second, called R2, has the energy $E=25111.8(2) \ {\rm cm}^{-1}$.
As one can see from Table~\ref{t:Fm}, the state R2 has anomalously small value of $A$. This means that the theoretical uncertainty for this state is large and the state is not very good for the extraction of nuclear parameters. In contrast, state R1 has relatively large values of $A$ and $B$ and therefore, present a better alternative for the analysis.



There is some difference in the results of the present work presented in Table~\ref{t:Fm} and the results of our previous calculations~\cite{ErFm}. This difference is due to some variation in the basis. It illustrates the accuracy of the method. This may lead to problems in identification of the states with close energies. For example, first two odd states with $J=6$ go in opposite order in the present and earlier calculations of Ref.~\cite{ErFm}. Therefore, it is important to know $g$-factors of the states as an additional mean of their identification. 
{ The good thing about $g$-factors is that they are more stable in the calculations. This is because they are proportional to the diagonal matrix element of the magnetic dipole transition (M1) operator which has no radial part. Therefore, the $g$-factors are not sensitive to the radial part of the wave function. On the other hand, they are sensitive to configuration mixing.}
Currently, no experimental data on $g$-factors are available. If measurements of the hfs are going to be used for extraction of nuclear parameters, then measuring $g$-factors becomes almost as important as measuring hfs itself. This is because wrong identification of the states may lead to wrong results for nuclear parameters.

\section{Conclusion}

We present calculations for energies, $g$-factors, and hfs constants $A$ and $B$ for 22 states of Fm.
Similar calculations for Er illustrate the accuracy of the applied method.
The results are to be used for the extraction of the nuclear magnetic dipole moments $\mu$ and nuclear electric quadrupole moments $Q$ from current and future measurements of the hfs in some Fm isotopes.

\acknowledgments

This work was supported  by the Australian Research Council  Grants No. DP230101058 and DP200100150.

\end{document}